\documentclass[aps,prl,twocolumn,superscriptaddress]{revtex4}  % for review and submission

\usepackage{graphicx}
\usepackage{amsmath}
\usepackage{enumerate,amsthm,amssymb,color}
\usepackage{dsfont}
\usepackage{bbm}
\usepackage{bm}
\usepackage{braket}
\usepackage{gensymb}
\usepackage{physics}
\usepackage[french, english]{babel}
\usepackage{float}
\usepackage{newfloat}
\usepackage{upgreek}
\usepackage{sidecap}
\begin{document}
\title{Active engineering of four-wave mixing spectral entanglement in hollow-core fibers}

\author{M. Cordier}\affiliation{Laboratoire de Traitement et Communication de l'Information, T\'el\'ecom ParisTech, Universit\'e Paris-Saclay, 75013 Paris, France}
\author{A. Orieux}\affiliation{Laboratoire d'Informatique de Paris 6, CNRS, Sorbonne Universit\'e, 75005 Paris, France}
\author{B. Debord}\affiliation{GPPMM Group, XLIM Research Institute, CNRS UMR 7252, Universit\'e de Limoges, Limoges, France} \affiliation{GLOphotonics S.A.S., 123 avenue Albert Thomas, Limoges, France}
\author{F. G\'erome}\affiliation{GPPMM Group, XLIM Research Institute, CNRS UMR 7252, Universit\'e de Limoges, Limoges, France} \affiliation{GLOphotonics S.A.S., 123 avenue Albert Thomas, Limoges, France}
\author{A. Gorse}\affiliation{GLOphotonics S.A.S., 123 avenue Albert Thomas, Limoges, France}
\author{M. Chafer}\affiliation{GPPMM Group, XLIM Research Institute, CNRS UMR 7252, Universit\'e de Limoges, Limoges, France} \affiliation{GLOphotonics S.A.S., 123 avenue Albert Thomas, Limoges, France}
\author{E. Diamanti}\affiliation{Laboratoire d'Informatique de Paris 6, CNRS, Sorbonne Universit\'e, 75005 Paris, France}
\author{P. Delaye}\affiliation{Laboratoire Charles Fabry, Institut d'Optique Graduate School, CNRS, Universit\'e Paris-Saclay, 91127 Palaiseau cedex, France}
\author{F. Benabid}\affiliation{GPPMM Group, XLIM Research Institute, CNRS UMR 7252, Universit\'e de Limoges, Limoges, France} \affiliation{GLOphotonics S.A.S., 123 avenue Albert Thomas, Limoges, France}
\author{I. Zaquine}\affiliation{Laboratoire de Traitement et Communication de l'Information, T\'el\'ecom ParisTech, Universit\'e Paris-Saclay, 75013 Paris, France}

%\affil[*]{Corresponding author: isabelle.zaquine@telecom-paristech.fr}
\date{\today}

%\ociscodes{(060.5295) Photonic crystal fibers; (190.4380) Nonlinear optics, four-wave mixing; (270.5565) Quantum communications.}
      
%\doi{\url{http://dx.doi.org/10.1364/ao.XX.XXXXXX}}

\begin{abstract}
We demonstrate theoretically and experimentally a high level of control of the four-wave mixing process in an inert gas filled inhibited-coupling guiding hollow-core photonic crystal fiber in order to generate photon pairs. The specific multiple-branch dispersion profile in such fibers allows both entangled and separable bi-photon states to be produced. By controlling the choice of gas, its pressure and the fiber length, we experimentally generate various joint spectral intensity profiles in a stimulated regime that is transferable to the spontaneous regime. The generated profiles cover both spectrally separable and entangled bi-photons and feature frequency tuning over 17 THz, demonstrating the large dynamic control offered by such a photon pair source.
\end{abstract}

%\begin{document}

\maketitle

\section{Introduction}
Entanglement is a key resource in quantum information and strong effort has been made to increase the Hilbert space dimension of the generated quantum states either through the number of particles involved or through the generation of qudits with dimension $d\gg2$ \cite{wang2016experimental, mair2001entanglement,krenn2014generation, malik2016multi,brecht2015photon}. However, in some applications, it is required to go against this trend and to look for separable states. This is particularly driven by the requirement to generate pure states, through heralding from multipartite states. One emblematic example is a heralded single photon source based on parametric photon pair generation. The purity of the heralded photon can be obtained only if the entanglement in all degrees of freedom (spatial, polarization, spectral) is removed between the photon and its heralding twin \cite{law2000continuous, grice2001eliminating}. In other terms, the photon-pair state must be engineered into a separable state. While removing spatial and polarization entanglement can be easily achieved, the suppression of spectral entanglement is generally challenging. It requires stringent conditions on the optical nonlinear process and more specifically on the phase-matching condition. This is exemplified with bulk crystals, commonly used in photon pair generation, and the limited wavelength range where the required conditions can be satisfied \cite{grice2001eliminating}. Integrated sources offer more flexibility in engineering the spectral entanglement at specific wavelengths, for instance by designing the poling period of the waveguide \cite{eckstein2011highly} or the microstructuration in photonic crystal fiber \cite{cohen_tailored_2009,soller2010bridging}. Within this context, tuning the photon pairs phase-matched frequencies has been reported using temperature control of a photonic chip \cite{kumar_controlling_2014}, or fiber \cite{ortiz2017spectral}. Furthermore, promising results have been obtained through gas pressure tuning using modulation instability in gas-filled hollow-core photonic crystal fibers, where the amount of spectral entanglement was controlled actively \cite{finger2017characterization}. However, this experiment was done in a high pump power regime and the results cannot be extrapolated to the photon-pair regime. \\
Here, we propose the first demonstration of spectral correlation engineering using multiband four-wave mixing process within a single photonic component, namely a gas-filled inhibited-coupling (IC) guiding hollow-core photonic crystal fiber (HCPCF) with controllable dispersion and optical nonlinearity. We used this feature to generate various joint spectral intensity (JSI) profiles, thus indicating production of various degrees of bi-photon entanglement.

\section{Spectral entanglement}

The bi-photon state produced by spontaneous four-wave mixing (SFWM) in an optical fiber of length \textit{L} is given below using a standard perturbative approach \cite{garay-palmett_photon_2007}:
\small
\begin{equation}
\ket{\psi_{pair}} = \kappa \int \int d\omega_s d\omega_i F(\omega_s,\omega_i) \, \hat{a}^{\dag}_s(\omega_s) \,\hat{b}^{\dag}_i(\omega_i) \ket{0,0}
\label{eq:hamiltonien}
\end{equation}
\normalsize
Here the quantity $\kappa$ is a normalizing constant, $\hat{a}^{\dag}_s$ and $\hat{b}^{\dag}_i$ are the monochromatic creation operators of the two photons $s$ and $i$ of the photon pair and  $F(\omega_s,\omega_i) $ is the joint spectral amplitude function (JSA) describing the spectral properties of the generated photon pair. This function can be approximated as the product of the energy conservation function $\alpha (\omega_s, \omega_i)$ and the phase matching function $\phi (\omega_s, \omega_i)$:
\small
\begin{equation}
F(\omega_s,\omega_i) \approx \alpha (\omega_s, \omega_i) \, \phi (\omega_s, \omega_i) 
\end{equation}
\normalsize
In FWM with degenerate pump, the energy conservation function is defined as the autoconvolution of the pump spectral amplitude  $\alpha = [A*A](\omega_s+\omega_i)$ with  $A(\omega)$ usually described by a Gaussian function of width $\sigma$.

The phase matching function is given by:
\small
\begin{equation}
\phi = \text{sinc}\Big[\Delta k_{\text{lin}}(\omega_s,\omega_i) \frac{L}{2}\Big]  \times \text{exp}\Big[ i \Delta k_{\text{lin}}(\omega_s,\omega_i) \frac{L}{2}\Big], 
\end{equation} 
\normalsize
where $\Delta k_\text{lin}$ is a first order Taylor expansion of the wavevector mismatch between the photon pair and the pump photon. It is performed around the perfect phase matched frequencies $\omega_s^0$, $\omega_i^0$ and $\omega_p^0$:
\small
\begin{align*}
\Delta k_\text{lin}(\omega_s,\omega_i) = \Delta k^{(0)}
&+ (\omega_s - \omega_{s}^0).(\beta_{1p}-\beta_{1s}) \\
&+ (\omega_i - \omega_{i}^0).(\beta_{1p}-\beta_{1i}),
\end{align*}
\normalsize
where $\beta_{1\mu} = \frac{dk}{d\omega}|_{\omega_\mu^0}$ is the inverse group velocity.
\begin{figure}[h]
\centering
\includegraphics[width=\linewidth]{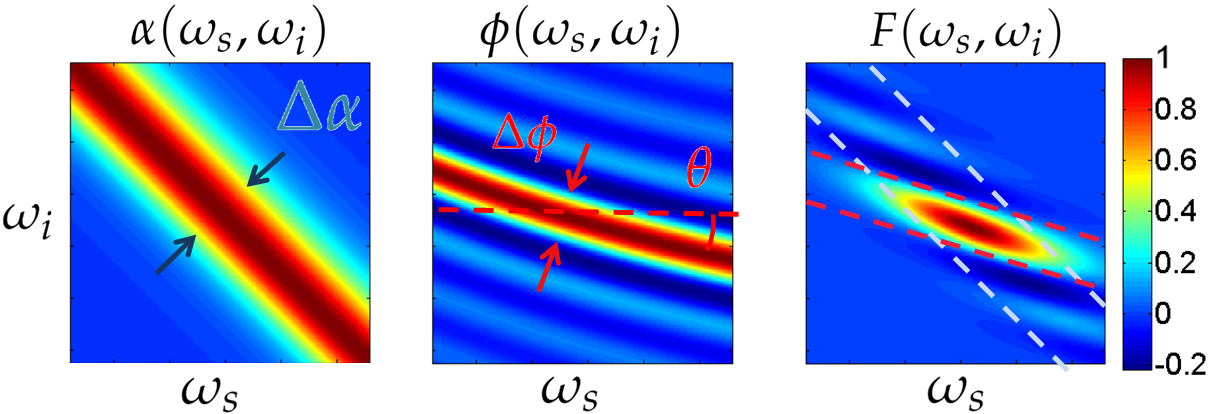}
\caption{\small From left to right: energy conservation function $\alpha$, phase matching function $\phi$ and corresponding joint spectral amplitude in the case of a Gaussian pump in an unengineered medium.} 
\label{fig:jsi}
\end{figure}

The spectral correlations can be displayed graphically by mapping $F$ in the ($\omega_s$, $\omega_i$) space as shown in Figure \ref{fig:jsi}.
The energy conservation function $\alpha$ exhibits spectral anti-correlation (Fig \ref{fig:jsi}, left). The shape of the phase matching function can either compensate or reinforce this correlation, depending on its width $\Delta \phi= |1/(2 L^2 (\beta_{1p}-\beta_{1s}). (\beta_{1p}-\beta_{1i}))|$ and angle $\theta= - \text{arctan} (\frac{\beta_{1p}-\beta_{1s} }{\beta_{1p}-\beta_{1i}})$.\\

\begin{figure}[h]
\centering
\includegraphics[width=\linewidth]{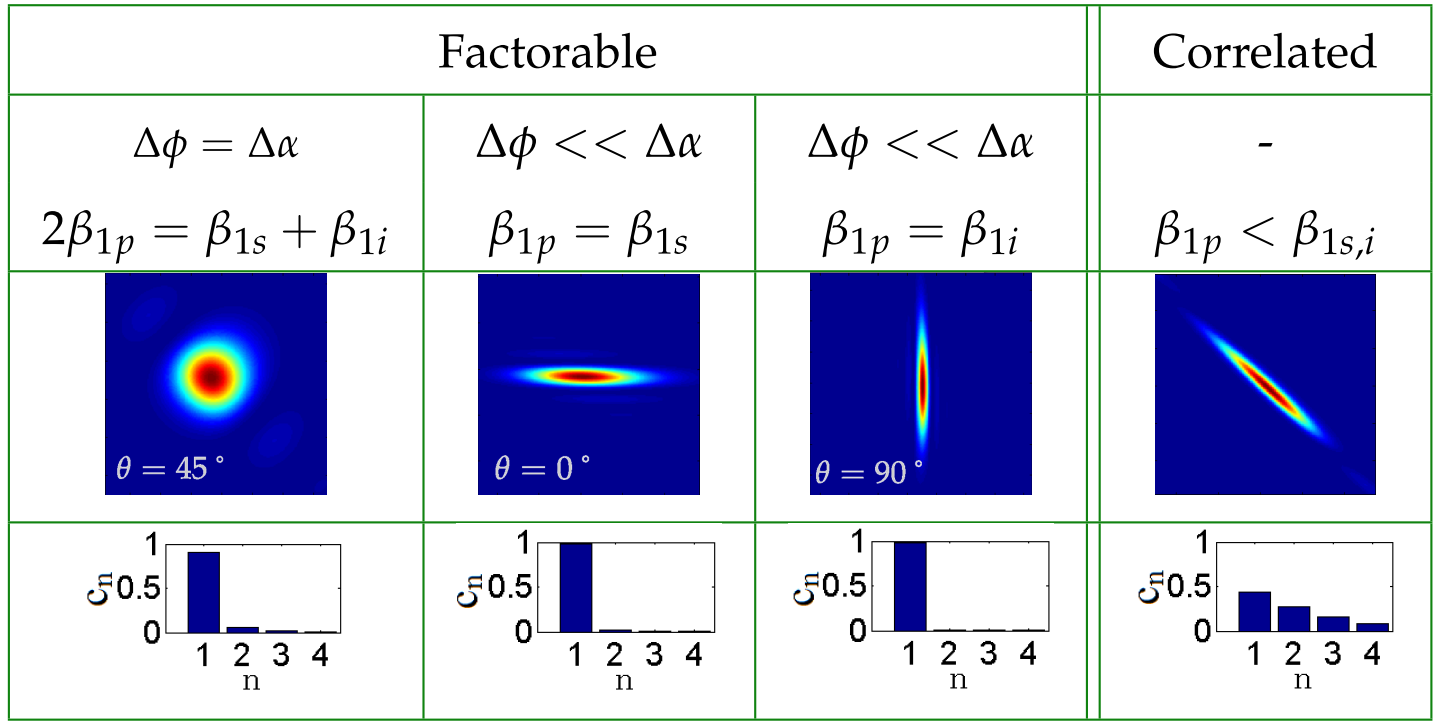}
\caption{\small Sets of conditions required to obtain either factorable or correlated pair states. Each set defines i) a relation between medium length and pump bandwidth ii) a relation between the group-velocities of pump, signal and idler photons which in turn determines the angle $\theta$.} 
\label{fig:tab}
\end{figure}

In fact, the JSA profile, chiefly described by $\theta$, $\Delta \alpha$ and $\Delta \phi$, is a direct signature of how the photon-pair states are correlated, as illustrated in Figure \ref{fig:tab} where we can distinguish two families. The first one comprises factorable states where the signal and idler photons, being spectrally independent, are in a product state. It has been shown \cite{grice2001eliminating} that this state can take the form of three distinctive JSA geometry profiles forming a circle or an ellipse along either the horizontal or the vertical axis, each corresponding to a specific phase relationship between the photon pair and the pump. The second family of states that can be identified by JSA profiles relates to what is commonly referred to as correlated states. In such a scenario signal and idler photons are spectrally entangled. A representative JSA of such states is shown on the right hand side panel of Figure \ref{fig:tab}. Here, the JSA typically shows a tilted ellipse profile. 

These various degrees of spectral entanglement can be described in a more quantitative way by using the Schmidt decomposition of the JSA, which consists in finding the two sets of orthonormal functions $S_n(\omega_s)$ and $I_n(\omega_i)$ depending only on the signal and idler frequency, respectively \cite{law2000continuous}: 
\small
\begin{equation}
F = \sum_{n=0}^\infty \sqrt{c_n} \, S_n(\omega_s) I_n(\omega_i) 
\end{equation} 
\normalsize
Substituting into Eq. \ref{eq:hamiltonien}, we obtain:
\small
\begin{equation}
\ket{\psi_{pair}} = \sum_{n=0}^\infty \sqrt{c_n} \, \hat{A}_n^{\dag} \, \hat{B}_n^{\dag} \ket{0,0},
\label{eq:Schmidt}
\end{equation}
\normalsize
where $\hat{A}_n^{\dag} = \int d\omega_s S_n(\omega_s) \hat{a}^{\dag}_s(\omega_s)  $ and $\hat{B}_n^{\dag} = \int d\omega_i I_n(\omega_s) \hat{b}^{\dag}_i(\omega_i)$ define the temporal modes creation operators. The coefficients \textbf{$c_n$} are real normalized scalars.

By definition, a factorable state corresponds to the case where there is only one non-zero element in the decomposition so that $\ket{\psi_{pair}} = \ket{A_0}\ket{B_0}$ implies spectral independence between signal and idler photons. We also define the mean number of excited modes by the Schmidt number : $K=\frac{1}{\sum_n c_n^2}$. $K=1$ for a factorable state and $K>1$ for a correlated state.

% \begin{table}[h!]
%  \centering \caption{Sets of conditions required to obtain either factorable or correlated pair states. Each set defines i) a relation between medium length and pump bandwidth ii) a relation between the group-velocities of pump, signal and idler photons which in turn determines the angle $\theta$.}
% \begin{tabular}{|c|c|c||c|}
% \hline
% \multicolumn{3}{|c||}{Factorable} & Correlated  \\
% \hline
% \small
% $\Delta \phi = \Delta \alpha $ & $\Delta \phi << \Delta \alpha $ & $\Delta \phi << \Delta \alpha $ & - \\
% $2\beta_{1p} = \beta_{1s} + \beta_{1i}$ & $\beta_{1p} = \beta_{1s}$ & $\beta_{1p} = \beta_{1i}$ & $\beta_{1p} < \beta_{1s,i}$ \\
% \hline
%  \includegraphics[width=\linewidth/2]{45final.PNG} & \includegraphics[width=\linewidth/2]{0final.PNG} & \includegraphics[width=\linewidth/2]{90final.PNG} & \includegraphics[width=\linewidth/2]{corre2.png}  \\
% % \hline
% % \includegraphics[width=\linewidth/5]{schmidt45.png} & \includegraphics[width=\linewidth/5]{schmidt0.png} & \includegraphics[width=\linewidth/5]{schmidt90.png} & \includegraphics[width=\linewidth/5]{schmidtcorre.png}\\
% % \hline
% \end{tabular}
%   \label{Table:JSA}
%     \end{table}

\normalsize

These photon-pair states, being spectrally factorable or entangled, are all useful in quantum technology (QT) applications. For example, correlated states are essential in security of quantum key distribution \cite{cerf2002security}, whilst factorable states are the backbone in heralded single photon sources. Furthermore, the temporal mode basis of the signal and idler presented above can be used as an encoding basis in a high dimension Hilbert space \cite{brecht2015photon}. For example, by controlling the Schmidt decomposition one can address the different qudit states of each photon of the pair and tailor the entanglement between these two qudits. A large range of control parameters is required in order to achieve that in a dynamic way with a given source, and we demonstrate here that hollow-core fibers provide the ideal platform for that purpose.

\section{Dispersion and group velocity matching conditions in Hollow-Core Fibers} \label{DHCF}
\begin{figure}[h!]
\centering
\includegraphics[width=\linewidth]{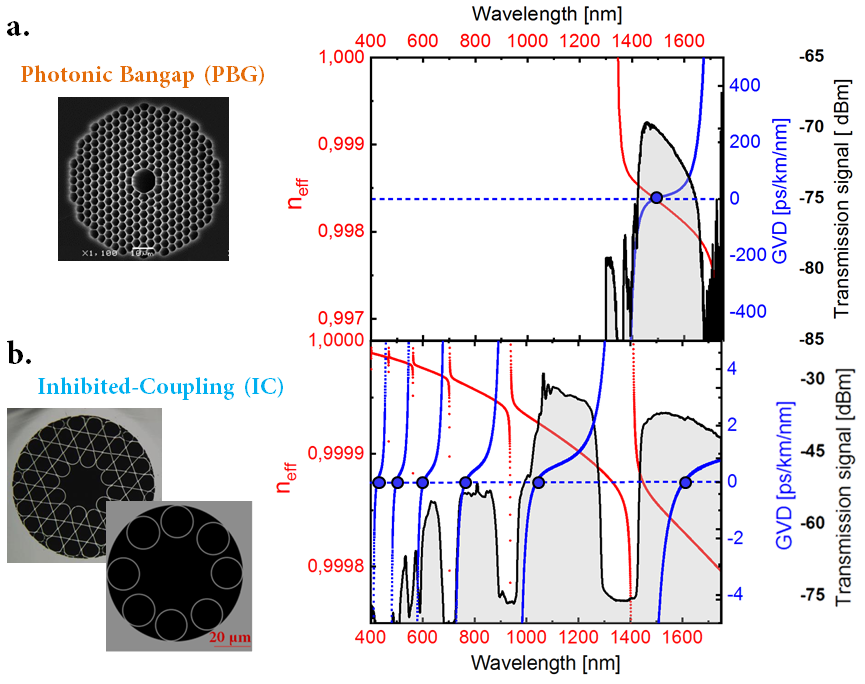}
\caption{\small \textbf{a.} PBG fiber and \textbf{b.} IC fiber cross sections and their optical properties. The effective index is plotted in red, the group-velocity dispersion in blue and the transmission in dark lines. The circles give the position of the zero-dispersion wavelengths.  }
\label{fig:fibers}
\end{figure}

HCPCFs are an exceptional tool for light and fluid-phase interactions thanks to their ability to micro-confine together light and fluids within small modal areas and over interaction lengths that can reach up to kilometers with extremely low optical losses. Furthermore, with a judicious choice of the fluid and HCPCF design, fluid-filled HCPCFs have been shown to be excellent and versatile platforms for nonlinear and quantum optics \cite{benabid2011linear,russell2014hollow}, whose optical nonlinearity and dispersion can be highly tailored. Today, we can distinguish two different types of HCPCFs whose typical transmission and dispersion spectra are illustrated in Figure \ref{fig:fibers}.

The first kind is photonic bandgap fiber (PBG) guiding HCPCF (PBG-HCPCF) \cite{birks1995full}. Here, the light guidance in the fiber core is achieved by engineering the 2D photonic crystal cladding whose modal spectrum at the effective indices and frequencies of the core-guided modes is void of any photonic state. The transmission spectrum often exhibits a single relatively narrow window. Consequently, the dispersion curve of the fundamental mode exhibits a ``tilted S shape" dispersion curve located within the spectral width of the transmission window (Fig. \ref{fig:fibers}.a, red curve), thus, exhibiting a single zero dispersion wavelength (ZDW). Such a dispersion profile generally limits the possible group-velocity relations that can be obtained. With the pump wavelength positioned near the ZDW and signal and idler photons generated on each side \cite{barbier2015spontaneous,Cordier:17}, $\beta_{1p}$ is near a minimum (defined by the position of the ZDW), so that  $\beta_{1p} < \beta_{1s}$ and $\beta_{1p} < \beta_{1i}$. Thus, only correlated states are generally accessible. Other group velocity matching conditions can be obtained in exotic PBG-HCPCF exhibiting a multiband dispersion profile \cite{cordier2017correlation}. However, even-though a double transmission band has been demonstrated \cite{light2009double}, multiband PBG-HCPCFs are usually achieved via anti-crossing between core modes and interface modes, which has practical drawbacks because of the hybrid nature of the involved spatial modes and of the difficultly in having a precise control on this multiband behavior.

The second type of HCPCF is Inhibited-Coupling guiding HCPCF (IC-HCPCF) \cite{couny2007generation}. IC-HCPCF does not require a PBG in its cladding structure, and its core guidance relies on a principle akin to bound or quasi-bound state in a continuum. In practice, this is achieved by engineering the fiber microstructure such that the coupling between the cladding mode continuum and the core mode is strongly inhibited. Figure \ref{fig:fibers}.b shows micrographs of two fiber designs of such IC-HCPCF, namely kagome-lattice \cite{benabid2002stimulated}, and tubular-amorphous lattice \cite{pryamikov2011demonstration,debord2017ultralow} and their typical transmission and dispersion spectra. 

\begin{table}[h!]
 \centering \caption{Features comparison between photonic bandgap and inhibited coupling HCPCF}
\begin{tabular}{|c|c|c|}
\hline
 & PBG & IC \\
 \hline
core diameter & 5-20 $\mu$m & 15-100 $\mu$m \\
\hline
strut thickness & 100-200 nm & 250-3000 nm \\
\hline
 loss& 1.2-25 @1550 nm & 10-50 @1550 nm \\
 $[$dB/km$]$& 50-100 @1 $\mu$m & 8.5-20 @1 $\mu$m \\
 & 1000 @500 nm & 10-20 @500 nm \\
\hline
bandwidth & 50-200 nm & $\sim$1 $\mu$m \\
\hline
 \end{tabular}
  \label{Table:comparo_fibre}
    \end{table}

Unlike PBG-HCPCF, IC-HCPCF intrinsically exhibit several bands in their transmission bandwidth spanning from the deep infrared to the UV (see Fig. \ref{fig:fibers}.b). The corresponding multiple ZDWs give rise to a large variety of phase-matching and group velocity relations.

An analytical model was recently given to describe the effective index of a $\text{HE}_{m,n}$ mode inside a tube-type hollow core fiber \cite{zeisberger_analytic_2017}:
\small
\begin{equation}
n_{\text{eff}} = n_{gas} - \frac{j_{m-1,n}^2}{2k_0^2n_{gas}R^2} - \frac{j_{m-1,n}^2}{k_0^3n_{gas}^2R^3}.\frac{\cot \big[\Psi(t)\big]}{\sqrt[]{\epsilon-1}} .\frac{\epsilon+1}{2},
\label{eq:model}
\end{equation}
\normalsize
with : $\Psi(t) = k_0 \, t \, \sqrt[]{n^2_{si} - n^2_{gas}} $, \textit{R} the fiber radius, $\epsilon = n^2_{si}/n^2_{gas}$, $j_{m,n}$ the $\text{n}^\text{th}$ root of the $\text{m}^\text{th}$ Bessell function, $t$ the silica strut thickness and $n_{si}$ the glass refractive index. Simulations and experimental observations have shown that such a tube-type model can provide an accurate description of IC-fibers, provided an effective fiber core radius $R_{eff}$ is used, that is larger than the actual one \cite{debordhypocycloid}. Moreover, as long as $\lambda<<R$ the evolution of the effective index is, in first approximation, independent of the cladding design around the core. IC-kagome and IC-tubular fibers can therefore exhibit very close dispersion profiles despite their apparent geometrical difference,  as long as they have the same radius and silica strut thickness. However, as losses strongly depend on the cladding geometry \cite{vincetti2016empirical}, this simple tube-type model is not suited to infer fiber losses.

The slowly varying contribution of the effective index is described by the first two terms of Eq. \ref{eq:model} depending on gas dispersion and fiber core radius. 
The third term introduces the resonances with the silica struts. The positions of these narrow non-guiding regions depend on silica strut thickness $t$ and correspond to $\Psi(t) =  j\pi$ with $j$ integer. This defines a set of wavelengths $\lambda_{j} = \frac{2t}{j} \sqrt{n_{si}^2 - n^2_{gas}}$ where the fiber dispersion is divergent.

\begin{figure}[h!]
\centering
\includegraphics[width=\linewidth]{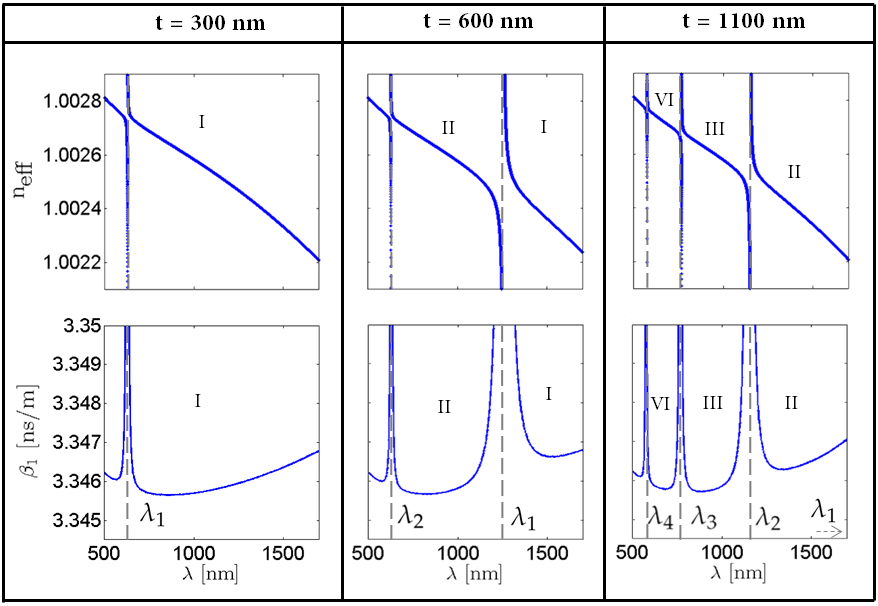}
\caption{\small Top: Effective index and bottom: inverse group velocity $\beta_1$ in a IC fiber filled with Xenon (4 bar) with a radius of 20 $\mu$m for three different silica strut thickness. The gray dashed lines correspond to the position of the discontinuities $\lambda_j$ and the Roman numbers to the different bands.}
\label{fig:IC}
\end{figure}

Figure \ref{fig:IC} shows the dispersion of IC-fibers calculated from Eq. \ref{eq:model} with different silica strut thicknesses $t$. For a small $t$ value, the discontinuities $\lambda_j$ are far from each other so that, for a given spectral window, the dispersion profile is mainly composed of one transmission band denoted I. As $t$ increases, the discontinuities get closer to each other, forming a multiband dispersion structure (additional bands denoted II, III, ...). The number and position of the bands and ZDWs can therefore be controlled when designing the fiber.

\begin{figure}[h!]
\centering
\includegraphics[width=0.85\linewidth]{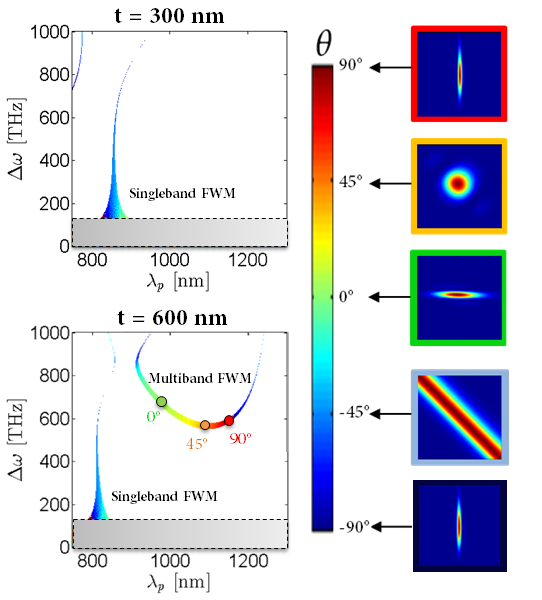}
\caption{\small FWM spectral density map for two different fiber geometries. The y-axis corresponds to the gap between pump and signal/idler frequencies $\Delta \omega = | \omega_p - \omega_{s/i} |$. The color gives an additional information about the angle $\theta$ of the phase matching function. The grey region corresponds to FWM where signal and idler are generated too close to the pump wavelength. Note that the lines have been thickened for improved visibility.}
\label{fig:pm_IC}
\end{figure}

Such multiband dispersion profiles provide an unprecedentedly large control parameter-space in setting the FWM wavelengths (i.e. pump, signal and idler). For example by positioning the three involved wavelengths either in the same band (singleband FWM) or in different bands (multiband FWM), one can control the phase relation between them to span the different bi-photon quantum states mentioned previously. This is illustrated in Figure \ref{fig:pm_IC} where we plot the FWM spectral density map giving both the wavelength of the generated photon and the angle $\theta$ of the phase matching as a function of the pump wavelength. For $t=300$ nm, the FWM $\lambda_p$, $\lambda_s$ and $\lambda_i$ are lying in the same band (band I). In such a case, $\theta$ varies from $-30\degree$ to $-80\degree$, thus corresponding to correlated JSA. 
For $t=600$ nm however, while this singleband configuration is still available, another one involving two bands is made possible where the idler is in band I while pump and signal wavelengths are in band II. When the fiber exhibits several bands, it is noteworthy that many relations between $\beta_{1p}, \beta_{1s}$ and $\beta_{1i}$ become possible allowing various angles $\theta$. More specifically, all three factorable states (i.e. $\theta = 0\degree$, $\theta = 45\degree$, $\theta = 90\degree$) become theoretically accessible. To our knowledge, this is the first demonstration that HC-PCF dispersion can be tailored to engineer such a large range of bi-photon quantum states. 

The core radius $R$ must also be chosen carefully as it is related to multiple properties, mainly: losses, generation efficiency and modal contents. The FWM generation efficiency varies as $R^{-4}$ which favors a small radius but the general losses also increase with decreasing core radius ($~R^{-4}$ for IC-tubular and $~R^{-3}$ for IC-kagome fiber \cite{vincetti2016empirical}) which forces a compromise. In such fibers, with a radius around tens of micrometers, the losses are remarkably low ($\approx 10-100$ dB/km) at a central wavelength in the band but they increase rapidly as the wavelength approaches a band-edge. None of the involved wavelengths $\lambda_p$, $\lambda_s$, $\lambda_i$ should therefore be too close to any discontinuity, which must be taken into account in the fiber design in order to enable efficient FWM. Finally, this type of fiber is usually a few-mode fiber but it can behave as single mode with a careful coupling.

\section{Tomography of the JSI in IC-HCPCF}

\begin{figure}[h!]
\centering
\includegraphics[width=\linewidth]{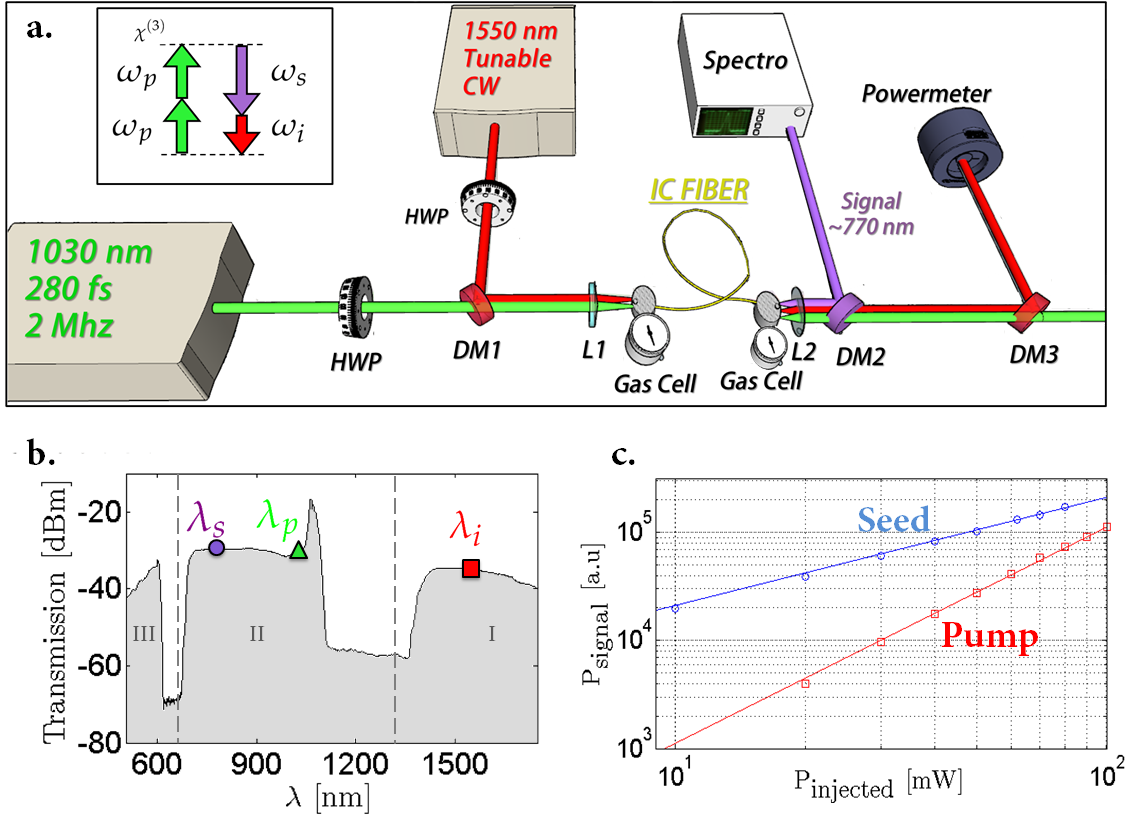}
\caption{\small \textbf{a.} Experimental setup for the stimulated emission tomography. HWP: half-wave plate, L: lens, DM: Dichroic Mirror. \textbf{b.} Measured fiber transmission for a 4.5 meter length fiber. The dashed lines give the position of the discontinuities predicted by the simulation. \textbf{c.} Power generated at the signal frequency as a function of the seed power (blue dots) and of the pump power (red squares), measured at the fiber output. The solid lines correspond to a perfect linear and quadratic dependence (log scale). }
\label{fig:manip}
\end{figure}

In order to test our model in practice we chose a tubular IC-HCPCF composed of 8 tubes with strut thickness $t = 625$ $\pm$ 20 nm. Its $22$ $\pm$ 1 $\mu$m effective core radius offers a good trade-off between losses, generation efficiency and modal content, as explained above. Cross-section and losses are given in Figure \ref{fig:fibers}.b. and Figure \ref{fig:manip}.b, respectively. The fiber design was chosen to exploit a multi-band FWM while operating at convenient wavelengths. Indeed, the pump wavelength is set at 1030 nm which is commercially very common. Similarly, the idler wavelength lies at telecom wavelength range, while the signal wavelength is in the range of atomic transitions and Silicon single photon detectors.Finally the fiber exhibits a polarization extinction ratio of $\sim$ 30 dB.

We use stimulated emission tomography technique (SET) to reconstruct the JSI ($= |\text{JSA}|^2$) \cite{liscidini2013stimulated}. Among the existing techniques, SET allows replacing the single photon detectors with fast and spectrally resolved classical detectors provided an additional laser is used at signal or idler wavelength to seed the process. The equivalence of the JSI between spontaneous and stimulated regime is valid only if the variation of the stimulated generated signal power remains linear with seed power \cite{fang2014fast, zielnicki2018joint}. The underlying assumption is that the JSI in the stimulated regime is an amplified version of the JSI in the spontaneous regime; the amplification factor is equal to the average photon number $N_{\text{seed}}$ of the coherent seed:  $\text{JSI}_{\text{stim}} \approx \, N_{\text{seed}}.\, \text{JSI}_{\text{spont}}$. 

Our SET experimental characterization setup is shown in Figure \ref{fig:manip}. 
Both pump and seed lasers are fibered. We use a pulsed pump laser (Satsuma, Amplitude Systems) at $\lambda_p =$ 1030 nm with 280 fs pulse duration at 2 MHz pulse repetition rate and the average pump power injected in the HCPCF is 60 mW. The seed laser (T100, Yenista) is a continuous-wave laser tunable from $\lambda_i$ = 1530 nm to 1560 nm. The seed power injected in the IC-HC-fiber is around 100 mW, which corresponds to a power of 50 nW  that is effectively involved in the FWM process because of the duty cycle of the pump. 
Both laser polarizations can be controlled independently with two halfwave plates (HWP). The two fiber ends are inserted into gas-cells with a pressure monitoring and glass windows for optical power injection. Injected into one of the cells, the gas fills the fiber core and an equilibrium between the two gas cells is obtained within less than an hour. At the fiber output, the generated signal at $\omega_s = 2\omega_p - \omega_{i}$ is filtered with dichroic mirrors and sent to a spectrometer (SILVER-Nova Stellar). For a given seed frequency $\omega_i$, the recorded spectrum gives one horizontal slice of the JSI. The full JSI is composed of all the recorded spectra obtained by sweeping the seed wavelength. The seed power is monitored in order to take into account any power change when tuning $\omega_i$ and to normalize the slices. In order to make sure that our measurements would be valid in the spontaneous regime, we have measured the power dependence of the power generated at signal frequency as a function of pump power and seed power. Figure \ref{fig:manip}.c shows the results confirming the linear dependence versus seed power and quadratic dependence versus pump power, thus validating that our experimental results are transferable to the spontaneous regime.

\begin{figure}[h!]
\begin{center}
\includegraphics[width=0.95\linewidth]{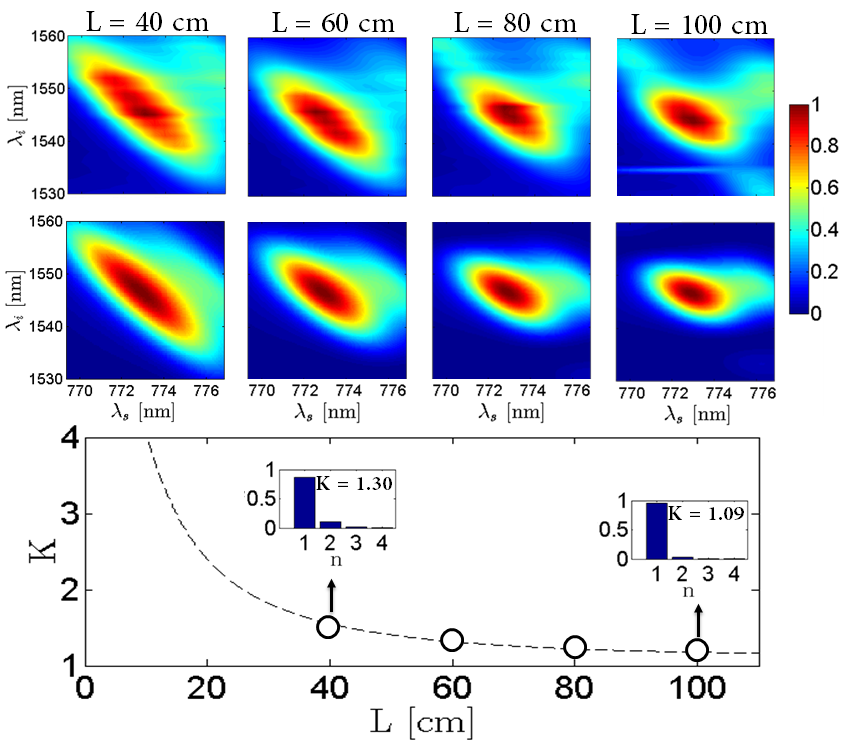}
\caption{\small Top, comparison between experimental (1st row) and simulated (2nd row) JSI for different fiber lengths when filled with 3.4 bar of Xe. The simulation takes into account a modulation in the Gaussian shape of the pump laser spectrum. Bottom, corresponding Schmidt number and Schmidt decomposition (assuming a flat phase). }
\label{fig2}
\end{center}
\end{figure}

Figure \ref{fig2} illustrates how, by simply changing the fiber length, the bi-photon JSI morphs from the shape of highly correlated states into that of nearly separable states. The figure shows a comparison between experimental and theoretically simulated JSI for fiber lengths going from 40 to 100 cm for the fiber filled with 3.4 bar of xenon as well as the associated Schmidt number. With 1 meter, the phase matching function and the energy conservation function have about the same widths, resulting in an almost factorable state. 
Decreasing the length implies an increase of the phase matching function width $\Delta \phi$, which gives an increasingly correlated JSI. The lobe-shaped region in the top right corner of the JSI profile is attributed to the non perfect Gaussian spectral shape of the pump laser pulses. Such shape is due to residual self-phase modulation taking place within the laser and which could be removed using appropriate filters or pulse shaping techniques. The parameters used in the simulation are $t = 630$ nm, $R_{eff} = 22$ $\mu$m and $P = 3.4$ bar. The radius is approximately 10\% higher than the actual fiber core measured with an optical microscope, as expected when extrapolating the tube-type dispersion model to that of IC fiber \cite{debordhypocycloid} (see section 2).

\section{Active tunability of the JSI}

\begin{figure*}[htp!]
\centering
\includegraphics[width=0.9\linewidth]{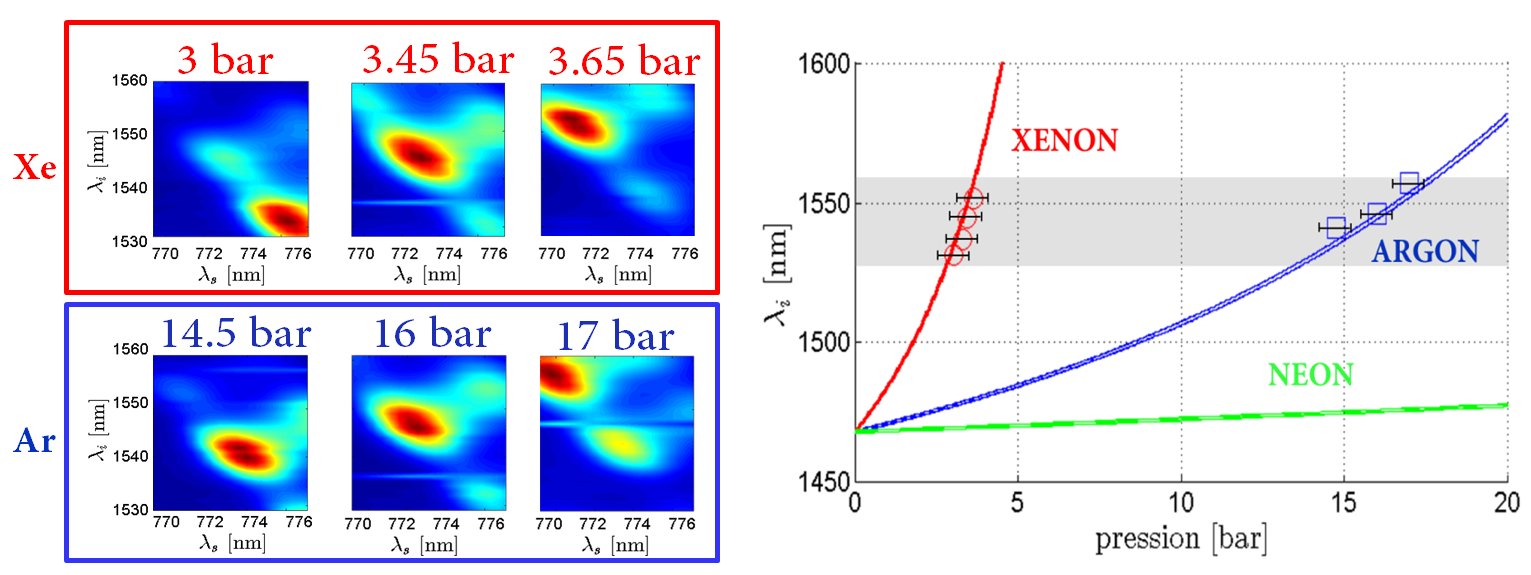}
\caption{\small Left: JSI as a function of gas pressure when filled with xenon and argon Right: Central position of the JSI idler wavelength as a function of gas pressure, for xenon (red circles) and argon (blue squares) and comparison with the simulation. The measurement range in shaded gray was limited by the tunability of the seed laser. }
\label{fig:pressure_pm}
\end{figure*}

An extended control of the spectral properties is possible through the filling gas. Gas temperature and, more notably, pressure can be used to actively impact the dispersion and nonlinear response.
The refractive index of the gas at temperature $T$ and pressure $P$ can be extrapolated from the general Sellmeir equation at standard temperature and pressure conditions ($T_0$,$P_0$)  using:  
\small
\begin{equation}
n_{\text{gaz}}(\lambda,P,T) \approx \sqrt[]{1 + (n_{\text{gaz}}^2(\lambda,P_0,T_0)-1).\frac{P}{P_0}.\frac{T_0}{T}}
\end{equation}
\normalsize
The overall dispersion in Eq. \ref{eq:model} results from a competition between the gas dispersion on one side, which is pressure- and temperature-dependent, and the guide dispersion on the other side. Thus, depending on the weight of one relative to the other, the overall dispersion is more or less sensitive to a change of gas refractive index. For instance, fibers with large core have a lower waveguide dispersion. Consequently they will be more sensitive to a change of gas refractive index, caused by a pressure modification. Alternatively, smaller core or fiber filled with light atomic weight gases, will be less sensitive to a change of gas refractive index as the waveguide dispersion will be predominant.  

Using the same fiber, we demonstrate this effect by characterizing the JSI of the source as a function of gas pressure, for two different gases, namely argon and xenon (see Fig. \ref{fig:pressure_pm}). Noble gases have been chosen because they are devoid of Raman scattering  \cite{azhar2013raman,lynch2013supercritical}, which eliminates the usual main source of noise in fibered photon-pair generation. The Sellmeier equations of xenon and argon are taken from \cite{hitachi2005new,bideau1981measurement} and their non-linear indices are $n_2 = 9.2$ $10^{-21} \, $ and 0.8 $10^{-21} \, \text{m}^{2} \text{/W/bar}$, respectively.

Increasing the pressure has two effects. Firstly, the signal and idler are generated further apart from the pump. For instance, we measured that changing the pressure from 3 bar to 3.65 bar of xenon shifts the idler wavelength from 1531 to 1551 nm and the signal wavelength from 776 to 771 nm. A linearization of the measured data gives a sensitivity of 27.4 Thz/bar in Xe and 5.6 Thz/bar in Ar, which can be compared with recent work on temperature tuning in solid core fibers where the reported sensitivity is 0.1 Thz/\degree C \cite{ortiz2017spectral}. Pressure tuning has the advantages of being compatible with long fibers and free of Raman noise as well as offering both high or low sensitivity according to the gas choice. Secondly, simulation also predicts that changing the pressure slightly rotates the angle $\theta$ of the phase matching of $\sim 6$\degree per bar of xenon. Adjusting $\theta$ can be used, for instance to fine-tune the purity of a heralded single photon source whereas wavelength tuning can be used for instance to aim at a specific telecom channel or atomic transition. With a tunable pump laser, it is also possible to tune the wavelength of one photon while keeping the angle constant and vice-versa.  

\section{Conclusion}
In this paper, we demonstrate that FWM in gas-filled hollow-core fibers provides a versatile platform for generating and manipulating photon-pair states. In particular we show how the intrinsic multiband dispersion of IC fibers can be exploited to access multiband FWM and therefore various JSI shapes corresponding to spectrally factorable or correlated photon pair states. In the case of inhibited-coupling fiber, we present a model relating fiber design, filling medium and dispersion properties to allow for the design of specific fibers for given quantum information applications. It is noteworthy that this very simple model involves only three design parameters, namely fiber core diameter, strut thickness and gas refractive index without adjustable parameter whatsoever.

The experimental results obtained with noble gases validate both the model and the concept of JSI manipulation. A near-factorable JSI is experimentally demonstrated with signal and idler at visible and telecom wavelengths respectively. The effect of fiber length variation on the degree of spectral entanglement as well as the dynamic tuning of the signal and idler wavelength over 17 THz through gas pressure are in perfect agreement with the simulations. Future work will investigate the tailoring of the pump spectral properties with a pulse-shaper for an improved control of the spectral entanglement. Furthermore, we believe this idea of using mode coupling to shape dispersion can be extended to other media such as waveguides and micro-ring resonators. 

\section*{Funding Information}
This work is supported by the "IDI 2016" project funded by the IDEX Paris-Saclay, ANR-11-IDEX-0003-02, Labex SigmaLim and funding from Region Nouvelle Aquitaine. 

\section*{Acknowledgments}
The authors would like to thank M. Raymer for helpful and insightful discussions.

% Bibliography
\bibliographystyle{apsrev}
\bibliography{sample}

\end{document}